\documentclass[letterpaper,american,aps, preprintnumbers, prb, reprint]{revtex4-1}
\usepackage[T1]{fontenc}
\usepackage[utf8]{inputenc}
\usepackage{float}
\usepackage{amsmath}
\usepackage{amssymb}
\usepackage{graphicx}



\usepackage{babel}
\begin{document}
\title{Internal spin resistance of spin batteries}
\author{Khuôn-Viêt \surname{PHAM}}
\email[]{khuon-viet.pham@universite-paris-saclay.fr}

\affiliation{Laboratoire de Physique des Solides, Université Paris-Saclay, 91405
Orsay Cedex, France}
\begin{abstract}
For spin batteries we introduce the concept of internal spin resistance
which quantifies the amount of backflow from the load to the battery.
It allows to relate through a Thévenin-Norton relation, spin current
sources to spin accumulation sources. The value of the internal spin
resistance is derived explicitly for several spin batteries based
on spin injection, ferromagnetic resonance or spin Hall effect. 
\end{abstract}
\maketitle

\section{Introduction. What is a spin battery? Internal spin resistance\label{sec:internal-spin-resistance}.}

Spin batteries are a crucial component of spintronics. The goal of
this paper is to argue that an important characteristic of spin batteries
is their internal spin resistance, which is the analog for spin batteries
of the internal resistance of charge batteries. Many specific realizations
of spin batteries have been proposed. The simplest spin current source
in spintronics is probably a ferromagnet through which a charge current
flows\citep{aronov_spin_1976,johnson_interfacial_1985}; a spin accumulation
and a spin current will appear in any adjacent paramagnetic metal:
this is the mechanism of \emph{spin injection} which is used in many
(so-called) non-local setups where a spin current flows separately
from a charge current. Spin pumping through ferromagnetic resonance\citep{brataas_spin_2002}
is another popular way to create a spin source. The \emph{spin Hall
effect}\citep{dyakonov_pis_1971,hirsch_spin_1999}, where a charge
current generates a transverse spin current can also be used for spin
current generation (e.g. Ref. \citep{kimura_room-temperature_2007}).
A solar spin battery\citep{zutic_proposal_2001} has been proposed
as well.

To the extent of our knowledge the idea of an internal spin resistance
has not been spelled out explicitly in the literature; the closest
would be in Ref. \citep{brataas_spin_2002}, where Brataas and co-workers
computed maximum spin voltage and maximum spin current generated by
the ferromagnetic resonance operated spin pump. As we will show later
the internal spin resistance for collinear spin batteries is just
the ratio of maximum spin voltage to maximum spin current.

Introducing the concept of internal spin resistance leads to distinguish
spin current batteries from spin accumulation sources; the obvious
practical import of that distinction is that this can help optimize
the design of a spin battery depending on whether one requires a strong
spin current (as in STT (spin transfer torque) applications) or strong
spin accumulations (as in GMR (giant magneto-resistance) ).

We will consider both collinear and non-collinear spin batteries and
in section \ref{sec:Illustrations.} will explicitly compute the internal
spin resistance for several spin batteries operated by spin accumulation,
spin Hall effect or ferromagnetic resonance.

\subsection{Defining internal spin resistance.}

\emph{We confine ourselves in this sub-section to collinear setups
so that all magnetizations have the same direction. The spin current
is also assumed to be one dimensional so that a single scalar can
describe it.} 

The simplest way to define a spin battery is as a spin current source.
But a spin current induces in general a non-equilibrium spin build-up:
therefore one could argue that a spin battery can also be viewed as
a spin accumulation source. This is quite obvious in Johnson-Silsbee
thermodynamic theory\citep{johnson_thermodynamic_1987} as well as
magnetoelectronic circuit theory\citep{brataas_finite-element_2000},
where spin accumulation is the counterpart of charge voltage and drives
both charge and spin currents.

When one turns to charge batteries, two alternative descriptions are
possible: battery as current source or as voltage source. The Thévenin-Norton
equivalence ensures that both points of view are interchangeable. 

When comparing charge batteries with spin batteries, spin accumulation
acts as the analog of (charge) voltage while spin current plays the
role of charge current. To make the analogy clearer we define a \textbf{spin
voltage} as 
\begin{align}
V_{s} & =-\left(\mu_{\uparrow}-\mu_{\downarrow}\right)/2e\label{eq:spinvolt}
\end{align}
(which is the spin accumulation cast in electrical units) and we will
also measure the spin current in electrical units:
\[
I_{s}=I_{\uparrow}-I_{\downarrow}\propto-e.
\]

If one connects a spin battery to a load (e.g. a paramagnetic metal)
a spin current will flow away; concurrently a spin accumulation will
build up in general in the load, inducing a backflow spin current
and spin accumulation in the battery. The spin current will therefore
be reduced in comparison to what it would be if there was no backflow.
Within linear-response this can be described through the equation:
\begin{equation}
I_{s}=I_{s,max}-\frac{V_{s}}{r_{IS}}\label{eq:1}
\end{equation}
 or
\begin{equation}
V_{s}=V_{s,max}-r_{IS}\:I_{s}\label{eq:2}
\end{equation}
 where $V_{s}$ is the spin voltage of the battery ($V_{s,max}$ being
its maximum value), $I_{s}$ its spin current (maximum $I_{s,max}$)
and $r_{IS}$ is the internal spin resistance of the battery. Obviously
exactly as with the internal resistance of a charge battery, 
\begin{equation}
r_{IS}=V_{s,max}/I_{s,max}
\end{equation}
 since for $V_{s}=0$, $I_{s}=I_{s,max}$ by Eq. (\ref{eq:1}), which
in turn implies by Eq. (\ref{eq:2}) $0=V_{s,max}-r_{IS}\:I_{s,max}$.

These relations imply a \emph{Thévenin-Norton equivalence between
spin current source and spin accumulation source }exactly as with
charge batteries\emph{. }The spin battery Thévenin-Norton relations
will be derived exactly for several examples of spin batteries in
Section \ref{sec:Illustrations.} in confirmation of the previous
simple argument.

One potentiel point of confusion is the fact that spin current and
spin accumulation are not uniform within the spin battery: which spin
currents and spin accumulations enter in the previous equations ?
Within a circuit theory, given the assumption of one-dimensional connections
between battery and load, the equations should be understood as \emph{local
relations at the point where spin current exits the battery to enter
the load}. Departures from one-dimensionnal spin current flow can
be handled through an averaging process so that the relations should
be quite general by replacing $I_{s}$ with $\left\langle I_{s}\right\rangle $
the average over the contact surface between battery and load, and
a similar definition for $\left\langle V_{s}\right\rangle $.

In case of interface discontinuities (for instance contact resistance)
spin current and spin voltage are understood to be defined \textbf{on
the battery side}.

Observe that the internal spin resistance depends solely on the spin
battery; it is by definition \emph{independent of the loads} connected
to the battery. The explicit values taken by $V_{s}$ and $I_{s}$
however depend on the load. 

\subsection{Ideal spin batteries\label{subsec:Ideal-spin-batteries.}.}

The definition of the internal spin resistance makes obvious two subsequent
definitions:
\begin{itemize}
\item ideal spin current source : $r_{IS}=\infty$;
\item ideal spin accumulation source: $r_{IS}=0$.
\end{itemize}
In terms of applications an ideal spin accumulation source should
be better suited to inducing large spin accumulations in the load,
which can be detected as charge voltages through Johnson-Silsbee charge-spin
coupling\citep{johnson_interfacial_1985,johnson_thermodynamic_1987}.
Ideal spin current sources should be more interesting when one is
in need of large angular momentum transfer such as with spin transfer
torques. 

In practice since one has always non-ideal sources, the relevance
of a spin battery to a given load should be measured by a comparison
of the internal spin resistance to the spin resistance of the load,
both being evaluated locally at the connection between battery and
load (the load spin resistance being the ratio $V_{s}^{load}/I_{s}$).

\subsection{Consequences on the response to a collinear spin battery\label{subsec:Consequences-of-spin}.}

Let us consider now a load connected to a spin battery. We will show
how the internal spin resistance modifies the response to a spin battery.

\begin{figure}

\noindent \begin{centering}
\includegraphics[width=0.8\columnwidth]{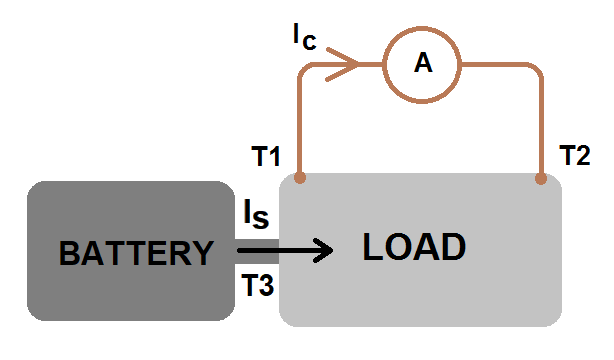}\caption{(color online) Three terminals are connected to the load. Terminal
$T_{3}$ links to the spin battery with incoming spin current $I_{s}$
(or spin accumulation $V_{s}$). The load response is measured across
terminals $T_{1}$ and $T_{2}$ as a charge current $I_{c}$ or a
voltage drop $V_{c}$.\label{fig:Three-terminals}}
\par\end{centering}
\end{figure}

\textbf{General response to a spin battery.} As noted by Brataas and
coll.\citep{brataas_spin_2002}, a spin battery is unipolar: there
needs be a single terminal connecting the battery and its load (while
a charge battery is of course dipolar). Through Johnson-Silsbee charge-spin
coupling a charge response of the load can be measured; this requires
at least two terminals through which either a voltage drop $V_{c}$
or a charge current $I_{c}$ is measured. The most general setup for
a spin battery driven device is therefore a three terminal one (see
Fig. \ref{fig:Three-terminals}). We call $T1$ and $T2$ the charge
measurement terminals on the load and $T3$ the terminal connecting
the spin battery to the load. 
 The charge response can be driven either by a spin current $I_{s}$
or a spin accumulation (or spin voltage $V_{s}$). If one assumes
there is no interface spin-flip, spin current is continuous at the
battery-load interface. 

Within linear response in the three terminal geometry of Fig. \ref{fig:Three-terminals}
this leads to linear relations :
\begin{align}
I_{s} & =g_{ss}\;V_{s}+g_{sc}\;V_{c},\label{eq:gss}\\
I_{c} & =g_{cs}\;V_{s}+g_{cc}\;V_{c}.\label{eq:gcs}
\end{align}
These transport parameters are load-dependent. Reciprocity implies
that $g_{cs}=g_{sc}$. If furthermore the two charge measurement terminals
are shorted, then $V_{c}=0$ and one measures $I_{c}$. Or conversely
a voltmeter can be used and then $I_{c}=0$. 

To have an idea of the magnitude of the transport coefficients of
the load, let us take a two channel model; then: 
\begin{align}
g_{cc} & =g_{ss}=g_{\uparrow}+g_{\downarrow}\label{eq:2ch}\\
g_{cs} & =g_{sc}=g_{\uparrow}-g_{\downarrow}=g_{cc}\;p
\end{align}
with $p$ being a conductance polarization. More generally one expects
therefore $g_{cc}$ and $g_{ss}$ to be of same order.

In addition to the 'charge-spin conductance'
\begin{equation}
g_{cs}=\frac{\partial I_{c}}{\partial V_{s}}=\left.\frac{I_{c}}{V_{s}}\right)_{V_{c}=0},
\end{equation}
several additional transport coefficients can also be defined to measure
the charge response to a spin drive:

\begin{equation}
r_{cs}=\left.\frac{\partial V_{c}}{\partial I_{s}}\right)_{I_{c}=0},\;h_{cs}=\left.\frac{\partial I_{c}}{\partial I_{s}}\right)_{V_{c}=0},\;k_{cs}=\left.\frac{\partial V_{c}}{\partial V_{s}}\right)_{I_{c}=0}
\end{equation}
 which have respectfully the dimensions of a resistance or are dimensionless.
The voltage drop $V_{c}$ and charge current $I_{c}$ are related
through the charge conductance $g_{cc}=\frac{\partial I_{c}}{\partial V_{c}}$
so that for instance $k_{cs}=-g_{cs}/g_{cc}$. While the charge conductance
$g_{cc}$ is always positive, $g_{cs}$ may be negative. 

It is against $g_{ss}$ that the internal spin resistance should be
compared as we shall now show. 

\textbf{Response to a Spin accumulation source.} The two relevant
transport coefficients are $g_{cs}$ and $k_{cs}$. 

Suppose one measures $I_{c}$ by shorting terminals $T1$ and $T2$
with an ammeter. Eq. \ref{eq:gss}-\ref{eq:gcs} apply with $V_{c}=0$.
We look for the transport coefficient: 
\begin{equation}
g_{cs}'=\frac{\partial I_{c}}{\partial V_{s,max}}
\end{equation}
 and want to show it will be reduced to the internal spin resistance.
For an ideal spin accumulation source the response would be $g_{cs}'=g_{cs}$
since $V_{s}=V_{s,max}$. 

The relation $V_{s}=V_{s,max}-r_{IS}\:I_{s}$ (eq. \ref{eq:1}) implies
$V_{s}=V_{s,max}-r_{IS}\:g_{ss}\:V_{s}$. Thus 
\begin{equation}
V_{s}=V_{s,max}/\left(1+r_{IS}\:g_{ss}\right).
\end{equation}
Therefore: 
\begin{equation}
g_{cs}'=\frac{g_{cs}}{\left(1+r_{IS}\:g_{ss}\right)}
\end{equation}

which means that the charge current $I_{c}$ measured will be reduced
by the factor 
\[
1+r_{IS}\:g_{ss}.
\]
 If we consider spin injection by a ferromagnet (see for instance
\ref{subsec:f-n} for an N-F-N trilayer acting as a spin battery)
$r_{IS}\sim r_{N}$ is in the $1\:\Omega$ range (if $r_{N}=\rho_{N}^{*}\:l_{N}/A$
is the spin resistance of the N layer, $l_{N}$ the spin relaxation
length in N, $A\sim10^{-15}m^{2}$ the cross-section for a nanopillar
for example). This means a load with $g_{ss}^{-1}\gg1\:\Omega$ is
better suited to that kind of battery. If the load is a paramagnetic
metal, the usual range is $g_{ss}^{-1}\sim1\:\Omega$.

The previous description assumes that there is no interface resistance
between the spin battery and the load. This is easy to incorporate
using:
\begin{equation}
V_{s}-V_{s}^{load}=r_{c}\:I_{s}
\end{equation}
which results in a further decrease of the response 
\begin{equation}
g_{cs}'=\frac{g_{cs}}{\left[1+\left(r_{c}+r_{IS}\right)\:g_{ss}\right]}.
\end{equation}
Inclusion of interfacial spin flip is straightforward and left to
the reader.

If one prefers to measure a voltage between terminals $T1$ and $T2$,
similar calculations lead to:
\begin{equation}
k_{cs}'=\frac{k_{cs}}{\left[1+r_{IS}\:\left(g_{ss}\;g_{cc}-g_{cs}\;g_{sc}\right)/g_{cc}\right]}
\end{equation}
 where 
\begin{equation}
k_{cs}'=\frac{\partial V_{c}}{\partial V_{s,max}}.
\end{equation}

In conclusion if $r_{IS}\ll g_{ss}^{-1}$ (typically this would mean
$r_{IS}\gg1\:\Omega$) the spin battery can be considered to be an
ideal spin accumulation source. 

\textbf{Response to a Spin current source.} The relevant transport
coefficients are now $r_{cs}$ and $h_{cs}$. 

Suppose first $V_{c}=0$ (terminals $T1$ and $T2$ shorted). The
relation $I_{s}=I_{s,max}-V_{s}/r_{IS}$ Eq. (\ref{eq:1}) combined
with Eq. (\ref{eq:gss}) implies $I_{s}=I_{s,max}-\left(r_{IS}\:g_{ss}\right)^{-1}\:I_{s}$.
Thus the incoming spin current is reduced
\[
I_{s}=I_{s,max}/\left(1+\left(r_{IS}\:g_{ss}\right)^{-1}\right).
\]
Therefore: 
\begin{equation}
h_{cs}'=\frac{I_{c}}{I_{s,max}}=\frac{h_{cs}}{\left(1+\left(r_{IS}\:g_{ss}\right)^{-1}\right)}
\end{equation}

which means that the charge current $I_{c}$ measured will be reduced
by the factor 
\[
1+\left(r_{IS}\:g_{ss}\right)^{-1}.
\]

If one rather measures a voltage between $T1$ and $T2$, similar
calculations lead to:
\begin{equation}
r_{cs}'=\frac{V_{c}}{I_{s,max}}=\frac{r_{cs}}{\left(1+\left[r_{IS}\:\left(g_{ss}\;g_{cc}-g_{cs}\;g_{sc}\right)/g_{cc}\right]^{-1}\right)}.
\end{equation}
If $r_{IS}\gg g_{ss}^{-1}$ there is no renormalization and the spin
battery can then be considered to be an ideal spin current source. 

\subsection{Non-collinear internal spin resistance: internal spin conductance
tensor.}

We keep the hypothesis of one-dimensional flow so that the spin current
is a vector instead of a rank-2 tensor and spin voltage is now a vector
instead of a scalar (see Eq. (\ref{eq:spinvolt}) ). For non-collinear
setups the obvious generalization of Eq. (\ref{eq:1}) is:
\begin{equation}
\mathbf{I_{s}}=\mathbf{I_{s,max}}-\underline{g_{IS}}\:\mathbf{V_{s}}
\end{equation}
or 
\begin{equation}
\mathbf{V_{s}}=\mathbf{V_{s,max}}-\underline{r_{IS}}\;\mathbf{I_{s}}
\end{equation}
where $\underline{g_{IS}}=\underline{r_{IS}}^{-1}$ is an internal
spin conductance tensor (see \ref{subsec:Bulk-spin-pumping.}-\ref{subsec:pumping}
for proofs of that relation for two examples of non-collinear spin
batteries). Eq. (\ref{eq:gss}-\ref{eq:gcs}) are replaced by:
\begin{align}
\mathbf{I_{s}} & =\underline{g_{ss}}\;\mathbf{V_{s}}+\mathbf{g_{sc}}\;V_{c},\\
I_{c} & =\mathbf{g_{cs}}\cdot\mathbf{V_{s}}+g_{cc}\;V_{c}.
\end{align}
where $\underline{g_{ss}}$ is a rank 2 tensor and the vectors $\mathbf{g_{cs}}$
and $\mathbf{g_{sc}}$ are equal by reciprocity. The charge current
response to a spin accumulation source is renormalized when one takes
into account internal spin resistance; instead of 
\[
I_{c}=\mathbf{g_{cs}}\cdot\mathbf{V_{s,max}}
\]
one has 
\begin{equation}
I_{c}=\left\{ \left[\left(1+\underline{r_{IS}}\;\underline{g_{ss}}\right)^{-1}\right]^{T}\mathbf{g_{cs}}\right\} \cdot\mathbf{V_{s,max}}.
\end{equation}
Similar expressions can be derived for the voltage $V_{c}$: 
\begin{equation}
V_{c}=\left\{ \left[\left(1+\underline{r_{IS}}\;\underline{g_{ss}'}\right)^{-1}\right]^{T}\mathbf{k_{cs}}\right\} \cdot\mathbf{V_{s,max}}
\end{equation}
where $V_{c}=\mathbf{k_{cs}}\cdot\mathbf{V_{s}}$ and 
\begin{equation}
g_{ss,ij}'=g_{ss,ij}-\frac{g_{cs,i}\;g_{cs,j}}{g_{cc}}.
\end{equation}

\subsection{Power dissipation.}

For collinear spin batteries there is complete isomorphism of internal
spin resistance to its charge counterpart since Eq. (\ref{eq:1})
has the same structure. We suppose spin current is exiting the battery
at a single point.

The spin battery is unipolar with $I_{c}=0$ but $I_{\uparrow}=-I_{\downarrow}$so
the spin battery is equivalent to a charge battery with the two poles
merged and with charge current $I_{\uparrow}$.

The power is : $\mathcal{P}=\sum_{\sigma}-I_{\sigma}\:\left(-\frac{\mu_{\sigma}}{e}\right)$
for current counted positive when leaving the device. Therefore: $\mathcal{P}=-V_{s}\:I_{s}$

One ends up with:
\begin{equation}
\mathcal{P}=-V_{s,max}\:I_{s}+r_{IS}\:I_{s}^{2}
\end{equation}
for a spin current in electrical units and counted positive when leaving
the battery. This expression shows that the internal spin resistance
is always positive.

For non-collinear spin batteries the power dissipated by the battery
is likewise:
\begin{equation}
\mathcal{P}=-\mathbf{V_{s}}\cdot\mathbf{I_{s}}=-\mathbf{V_{s,max}}\cdot\mathbf{I_{s}}+\mathbf{I_{s}}\cdot\underline{r_{IS}}\:\mathbf{I_{s}}
\end{equation}
which implies that $\underline{r_{IS}}+\underline{r_{IS}}^{T}$ is
a positive semi-definite matrix to ensure correct sign for power dissipation.

\section{Illustrations\label{sec:Illustrations.}.}

We explicitly derive the internal spin resistance for five spin batteries:
(i) a single ferromagnetic layer which induces spin accumulation in
neighbouring paramagnetic metals through spin injection , (ii) a F
- N - F trilayer, (iii) bulk spin pumping, (iv) the ferromagnetic
resonance operated spin battery of Brataas and coworkers, (v) spin
Hall effect.

\subsection{Single ferromagnetic layer \label{subsec:f-n}.}

The prototypal spin battery is a single ferromagnetic layer through
which a charge current flows. This is a spin accumulation and spin
current source for paramagnets thanks to spin injection. We now compute
its characteristic spin battery parameters, namely its internal spin
resistance and its maximum spin voltage.

\begin{figure}[H]
\begin{centering}
\includegraphics[width=0.8\columnwidth]{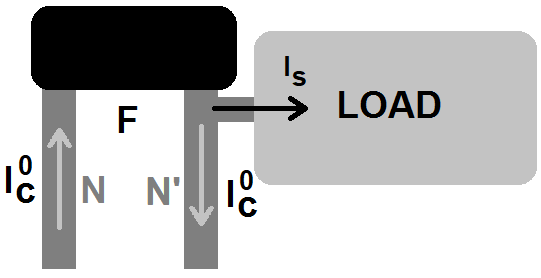}
\par\end{centering}
\caption{\label{fig:bat1}N - F - N trilayer used as a spin emf battery. The
load should be connected close to an interface for maximum spin voltage.
The charge current $I_{c}^{0}$ crosses the ferromagnet $F$ and induces
through spin injection a spin current $I_{S}$ in the load as well
as a spin accumulation $V_{s}$.}
\end{figure}

We consider the following non-local geometry: a paramagnetic N electrode
brings current to a ferromagnetic F rod connected to a second paramagnetic
N' electrode which collects the current. For maximum spin voltage,
the load must be connected close to an interface. In our model we
have chosen the load connection at the interface F-N' on the N' side
but it could have been on the F side (see Fig.\ref{fig:bat1}). 

We rely on a one-dimensional modellization within the standard drift-diffusion
formalism\citep{johnson_thermodynamic_1987,valet_theory_1993,rashba_diffusion_2002}.
The ferromagnet has length $d$ and is sandwiched betwwen two paramagnetic
metals N and N' which are assumed to be infinite. The origin $O$
is in the middle of F and the F-N interfaces are located at $O_{L}\left(z=-d/2\right)$
and $O_{R}\left(z=d/2\right)$. The spin current at $O_{L}$ is continuous
$I_{s}^{L}=I_{s}\left(O_{L}\right)$ if one neglects interfacial spin-flip;
at the other interface $O_{R}$ however:
\begin{equation}
I_{s}^{R}=I_{s}^{F}\left(O_{R}\right)=I_{s}^{N}\left(O_{R}\right)+I_{s}^{load}
\end{equation}
which expresses the spin current continuity. All the spin currents
are counted positive in the direction of axis $z$ (or away from the
battery for $I_{s}^{load}$). 
\begin{figure}[H]
\begin{centering}
\includegraphics[width=0.8\columnwidth]{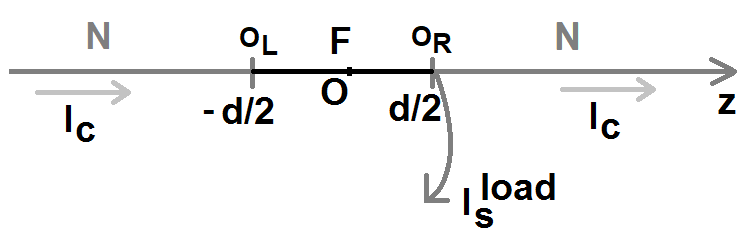}
\par\end{centering}
\caption{\label{fig:bat1geo}Ferromagnet used as a spin battery through spin
injection. The load should be connected close to an interface for
maximum spin voltage.}
\end{figure}
 The paramagnets are supposed to be infinite. After tedious but straightforward
calculations one eventually gets 
\begin{eqnarray}
r_{IS} & = & r_{N}-\frac{r_{N}^{2}}{r_{F}}\:\frac{1+X_{1}\:\tanh\frac{d}{l_{F}}}{\left(X_{1}+X_{2}\right)+\left(1+X_{1}X_{2}\right)\,\tanh\frac{d}{l_{F}}},\\
X_{i} & = & \frac{r_{N}+r_{ci}}{r_{F}},\;i=1,2
\end{eqnarray}
($l_{F}$ is the ferromagnet spin relaxation length , $d$ is the
ferromagnet length, $l_{N}$ is the paramagnet spin relaxation length,
$r_{N}=\rho_{N}^{*}\:l_{N}/A$ is the paramagnet spin resistance with
$A$ its cross-section, $r_{ci}$ ($i=1,2$) are the contact resistances
at the interfaces $O_{L}$ and $O_{R}$). 

We can simplify the expressions by assuming identical interface parameters
at $O_{L}$ and $O_{R}$, $P_{c1}=P_{c2}=P_{c}$ (interface conductance
polarizations, noted usually $\gamma$ in Valet-Fert notations\citep{valet_theory_1993})
and $r_{c1}=r_{c2}=r_{c}$, then with $X=\left(r_{N}+r_{c}\right)/r_{F}$:

\begin{equation}
r_{IS}=r_{N}-\frac{r_{N}^{2}}{r_{F}}\:\frac{1+X\:\tanh\frac{d}{l_{F}}}{2X+\left(1+X^{2}\right)\,\tanh\frac{d}{l_{F}}},
\end{equation}

and the maximum spin voltage is found as: 
\begin{equation}
V_{s,max}=r_{N}\,I_{c}\:\left[\frac{P_{c}\,r_{c}+P_{F}\,r_{F}\,\tanh\frac{d}{2l_{F}}}{r_{N}+r_{c}+r_{F}\,\tanh\frac{d}{2l_{F}}}\right]
\end{equation}

(where $P_{F}$ is the usual conductivity polarization of the ferromagnet,
noted $\beta$ in Valet-Fert notation\citep{valet_theory_1993}).
The term in brackets is actually the current polarization ($I_{s}/I_{c}$);
it is bounded by $\left[P_{c}\,r_{c}+P_{F}\,r_{F}\,\tanh\frac{d}{2l_{F}}\right]/\left[r_{c}+r_{F}\,\tanh\frac{d}{2l_{F}}\right]$
which is a weighted average of $P_{c}$ and $P_{F}$. Therefore the
larger the spin injection, the larger the spin voltage.

Let us examine limiting cases:
\begin{itemize}
\item $d\rightarrow0$ (thin ferromagnet): then the spin voltage is dominated
by interfaces and one gets:
\begin{align*}
V_{s,max} & \rightarrow\frac{r_{N}\,r_{c}\,P_{c}}{\left(r_{N}+r_{c}\right)}\,I_{c};\\
r_{IS} & \rightarrow\frac{r_{N}\:r_{c}}{r_{N}+r_{c}}.
\end{align*}
 This implies the spin voltage is set by the smallest of either the
contact or normal layer resistance. For Co-Cu layers for instance,
using $P_{c}\sim0.75$, $P_{F}\sim0.5$, $r_{F}\sim10\:\Omega$, $r_{N}\sim1\:\Omega$,
$r_{c}\sim1\:\Omega$ (ranges extracted from Ref. \citep{valet_theory_1993}
and \citep{bass_spin-diffusion_2007}) with an area $A=1\:fm^{2}$and
$I_{c}\sim0.1\:mA$ yielding $V_{s,max}\sim0.01\:mV=10\:\mu V$ and
$r_{IS}\sim0.5\:\Omega$.
\item $d\rightarrow\infty$ (thick ferromagnet): then the spin voltage becomes
\begin{align*}
V_{s,max} & \rightarrow r_{N}\,I_{c}\,\left[\frac{P_{c}\,r_{c}+P_{F}\,r_{F}}{r_{N}+r_{c}+r_{F}}\right];\\
r_{IS} & \rightarrow\frac{r_{N}\:\left(r_{c}+r_{F}\right)}{r_{N}+r_{c}+r_{F}}.
\end{align*}
This is also in the $10\:\mu V$ range for $V_{s,max}$ while $r_{IS}\sim1\:\Omega$.
More generally, the scale of $V_{s,max}$ is set by $r_{N}\:I_{c}\sim10\:\mu V$
so we don't expect to go much beyond the $10\:\mu V$ range with this
kind of battery.
\end{itemize}
The internal spin resistance scale is given by $r_{N}$ which is its
maximum value. It is therefore in the $1\:\Omega$ range.

It is interesting to compute a conversion gain $\mathcal{C}_{cs}=V_{s}/V_{c}=V_{s}/\mathcal{R}\,I_{c}$
which is a measure of the efficiency to convert a charge voltage into
a spin voltage. One finds: 
\begin{equation}
\mathcal{C}_{cs}=\frac{r_{N}}{\mathcal{R}}\,\left[\frac{P_{c}\:r_{c}+P_{F}\:r_{F}\:\tanh\frac{d}{2l_{F}}}{r_{N}+r_{c}+r_{F}\,\tanh\frac{d}{2l_{F}}}\right].
\end{equation}
For metallic multilayers, the resistance of the trilayer N-F-N' $\mathcal{R}$
is in the $10\:\Omega$ range; plugging in the figures we used for
a Co-Cu interface, this yields $\mathcal{C}_{cs}\sim1\%$ which is
small and consistent with the figures $V_{s,max}\sim10\:\mu V$ (for
$V_{c}\sim1\:mV$).

\subsection{F-N-F trilayer \label{subsec:f-n-f}.}

\begin{figure}[H]
\begin{centering}
\includegraphics[width=0.8\columnwidth]{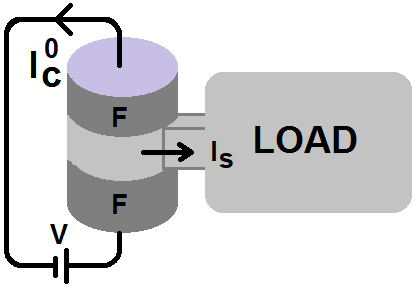}
\par\end{centering}
\caption{(color online)\label{fig:fnf battery}F - N - F nanopillar used as
spin battery. The load is connected to the spin battery through the
N paramagnetic layer. A charge current $I_{c}^{0}$ crossing the nanopillar
ensures spin injection occurs into the load (as a spin current $I_{s}$
or a spin voltage $V_{s}$). }
\end{figure}
\begin{figure}[H]
\begin{centering}
\includegraphics[width=0.8\columnwidth]{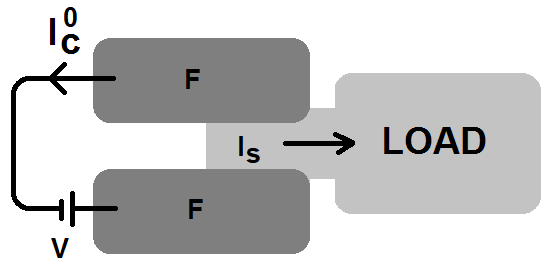}
\par\end{centering}
\caption{\label{fig:fnf battery2}Non-local geometry for F-N-F spin battery.
The load is connected to the spin battery through the N paramagnetic
layer. }
\end{figure}

We consider now the following spin battery which will prove more efficient
than the previous one: a F - N - F trilayer through which flows a
charge current. A spin accumulation is generated in the N layer which
is wired to a load. In the latter flows a spin current. Fig.\ref{fig:fnf battery}
shows a nanopillar geometry while Fig.\ref{fig:fnf battery2} is a
non-local variant. 

\begin{figure}[H]
\noindent \begin{centering}
\includegraphics[width=0.8\columnwidth]{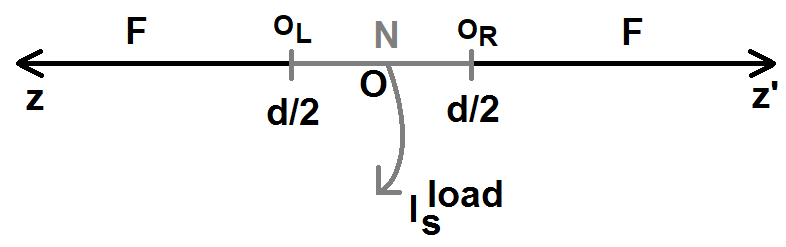}\caption{F-N-F spin battery. The load is connected to the paramagnetic N layer
(a spin current $I_{s}^{load}$ flows out of the battery to the load).
On the left and the right of origin $O$ we use different axes $z$
and $z'$.\label{fig:F-N-F-spin-battery.} }
\par\end{centering}
\end{figure}

We model the spin battery in one dimension using the standard drift-diffusion
equations\citep{johnson_thermodynamic_1987,valet_theory_1993,rashba_diffusion_2002}.
The origin is in the middle of the N layer. It is convenient to use
opposite axes to the left ($z$) and to the right $(z'$) of $O$.
Charge current $I_{c}$ flows from left to right. The width of the
N layer is $d$ and the interfaces with the ferromagnets are located
at $O_{L}\left(z=d/2\right)$ and $O_{R}\left(z'=d/2\right)$ (see
Fig.\ref{fig:F-N-F-spin-battery.}). The spin currents at the interfaces
$I_{s}^{L}=I_{s}\left(O_{L}\right)$ and $I_{s}^{R}=I_{s}\left(O_{R}\right)$
are continuous if one neglects interfacial spin-flip; they are counted
positive in the direction of axes $z$ and $z'$ respectively. 
The ferromagnets are supposed to be infinite. We use the same notations
as in previous section. After tedious but straightforward calculations
one can show that: 
\begin{eqnarray}
r_{IS} & = & \frac{2\,r_{N}\:\left[\sinh\frac{d}{2l_{N}}+X_{i}\:\cosh\frac{d}{2l_{N}}\right]}{\Delta},\\
X_{i} & = & \frac{r_{Fi}+r_{ci}}{r_{N}},\nonumber \\
\Delta & = & \exp\frac{d}{l_{N}}\,\prod_{i=1,2}\left(1+X_{i}\right)-\exp-\frac{d}{l_{N}}\,\prod_{i=1,2}\left(1-X_{i}\right)\nonumber 
\end{eqnarray}
($l_{N}$ is the paramagnet spin relaxation length) and the maximum
spin voltage is:

\begin{eqnarray}
V_{s,max}=\\
\frac{2\,I_{c}}{\Delta}\:\left\{ \left[P_{c2}\,r_{c2}+P_{F2}\,r_{F2}\right]\right. & \left[\sinh\frac{d}{2l_{N}}+X_{1}\:\cosh\frac{d}{2l_{N}}\right]\nonumber \\
-\left[P_{c1}\,r_{c1}+P_{F1}\,r_{F1}\right] & \left.\left[\sinh\frac{d}{2l_{N}}+X_{2}\:\cosh\frac{d}{2l_{N}}\right]\right\} .\nonumber 
\end{eqnarray}

This implies that the spin voltage is larger for antiparallel ferromagnetic
electrodes: $P_{F1}=-P_{F2}$ and $P_{c1}=-P_{c2}$ while it vanishes
for parallel magnetizations.

In the limit of thin width for the N layer ($d\ll l_{N}$), 
\begin{eqnarray}
r_{IS} & \longrightarrow & \frac{\left(r_{F1}+r_{c1}\right)\,\left(r_{F2}+r_{c2}\right)}{r_{F1}+r_{c1}+r_{F2}+r_{c2}}
\end{eqnarray}

For identical ferromagnets the internal resistance scales as $r_{I,s}=\left(r_{F}+r_{c}\right)/2$. 

In the same limit ($d\ll l_{N}$), the spin voltage for identical
antiparallel ferromagnet becomes:
\begin{equation}
V_{s,max}=I_{c}\:\left[P_{c}\,r_{c}+P_{F}\,r_{F}\right].
\end{equation}

Quantitatively for a Co - Cu - Co trilayer with antiparallel ferromagnets,
using\citep{bass_spin-diffusion_2007,valet_theory_1993} $P_{c}\sim0.75$,
$P_{F}\sim0.5$, $r_{F}\sim10\:\Omega$, $r_{c}\sim1\:\Omega$ with
an area $A=1\:fm^{2}$and $I_{c}=0.1\:mA$ yields, $V_{s,max}\sim1\:mV$
for a conversion gain $\mathcal{C}_{cs}=V_{s}/V_{c}$ which will be
close to unity $\mathcal{C}_{cs}\sim1$ which is much better than
the 1\% we found for the simple spin battery of Section \ref{subsec:f-n}.
The spin voltage is therefore potentially much larger than for the
spin battery made out of a single ferromagnet. The physical reason
behind that is simple: it stems from the much smaller spin relaxation
volume within the paramagnet.

For the same parameters one gets $r_{IS}\sim10\:\Omega$ while typically
$g_{ss}^{-1}\sim g_{c}^{-1}\sim1\:\Omega$; this F-N-F trilayer is
better as a spin current source for a paramagnetic metallic load.
If one looks for a better spin accumulation source, loads with smaller
$g_{ss}$ need to be picked. 

\subsection{Bulk spin pumping\label{subsec:Bulk-spin-pumping.}.}

Let us illustrate the computation of the internal spin resistance
in a non-collinear setting. The calculations are quite simple in a
spin pump model proposed by Watts and co-workers\citep{watts_unified_2006}.
We are not aware of any experimental implementation so we use this
spin pump as a toy model and refer the reader to the original publication
for a discussion of the physical relevance. The spin pumping relies
on an rf rotating magnetic field which generates a dc spin accumulation
within non-magnetic materials; this is a bulk mechanism in contrast
with Brataas and co-workers spin battery operated by ferromagnetic
resonance\citep{brataas_spin_2002} and governed by interface effects
between the ferromagnet and a paramagnetic load (i.e. the spin mixing
conductance at the interface). The bulk spin pump model is in principle
applicable to metals and semiconductors.

The basis of this model is a semi-classical equation for the non-equilibrium
spin accumulation vector $\vec{f}$ :
\begin{equation}
-\partial_{t}\vec{f}-g\:\mu_{B}\:\dot{\vec{B}}=\frac{\vec{f}}{\tau}-D\:\nabla^{2}\vec{f}+\frac{g\:\mu_{B}}{\hbar}\:\vec{B}\times\vec{f}
\end{equation}

The novel feature in this equation is the magnetic field time derivative
which is described as a Zeeman shift of the chemical potential for
each spin species. The space is divided in two metallic paramagnetic
regions: for $x<0$ region $(I)$ where there is a rotating magnetic
field 
\begin{equation}
\vec{B}=\left(\begin{array}{ccc}
B_{xy}\,\cos\omega t, & B_{xy}\,\sin\omega t, & B_{z}\end{array}\right)
\end{equation}
 and for $x>0$ region$(II)$ where $\vec{B}=0$. A solution which
is stationary in the rotating frame is readily found in region $I$
:
\begin{equation}
\vec{f}(x,\,t)=\vec{f}_{uni}+\alpha\:\left(\vec{\omega}-\vec{\omega_{B}}\right)\tau\:\exp\left(x/l_{sf}\right)
\end{equation}
where $\vec{\omega}=\omega\,\hat{z}$, $\vec{\omega_{B}}=\frac{g\:\mu_{B}}{\hbar}\:\vec{B}$,
$l_{sf}$ is the spin diffusion length and $\vec{f_{uni}}$ is the
uniform solution found when the rotating field is applied in the whole
space. $\alpha$ is a constant determined by boundary conditions.

The spin current density in electrical units is :
\begin{equation}
\vec{j_{S}}(x)=\frac{1}{\rho_{N}^{\text{*}}}\:\partial_{x}\vec{f}(x)
\end{equation}

Therefore the spin current $\mathbf{I_{s}}=\vec{j_{S}}\:A$ (where
$A$ is the cross-section) is : 
\begin{equation}
\mathbf{I_{s}}(0)=\frac{1}{r_{N}}\:\left(\vec{f}(0)-\vec{f_{uni}}\right)
\end{equation}
where the spin resistance is $r_{N}=\rho_{N}^{\text{*}}\:l_{sf}/A$.
This implies the following relation between the spin current $\mathbf{I_{s}}$
and the spin voltage $\mathbf{V_{s}}=-\vec{f}(0)$ : 
\[
\mathbf{I_{s}}=\mathbf{I_{s,0}}-g_{IS}\:\mathbf{V_{s}}
\]
 with a diagonal internal spin conductance tensor : 
\begin{equation}
g_{IS}^{ij}=\frac{\delta_{ij}}{r_{N}}.
\end{equation}
The internal spin resistance is therefore in the $1\:\Omega$ range
for paramagnetic metals in this kind of spin battery. 

\subsection{Interface spin pumping \label{subsec:pumping}.}

The ferromagnetic resonance operated spin battery\citep{brataas_spin_2002}
requires a non-vanishing spin mixing conductance at a F-N interface.
It is an interface driven spin pump. The defining equation for the
internal spin resistance eq. (\ref{eq:1}) must therefore be understood
to be evaluated on the N side of the F-N interface (in other words
it is crucial to include the interface within the spin battery).

Two regimes for the ferromagnetic operated spin pump can be distinguished\citep{brataas_spin_2002,brataas_non-collinear_2006,wang_voltage_2006}
depending on whether the load is a good or a bad spin sink (i.e. whether
the spin relaxation time in the load is small or large); the good
spin sink regime corresponds to a regime where Gilbert damping is
enhanced while in the bad spin sink case the spin pump is said to
be in the spin battery regime. As will be shown shortly the bad spin
sink regime corresponds to the spin pump being a good spin accumulation
source while the good spin sink regime corresponds to a good spin
current source. 

It can be shown that in the limit of vanishing spin flip in the F
layer, the longitudinal spin current vanishes\citep{wang_voltage_2006}.
The total spin current (in electrical units) is then reduced to:
\begin{align}
\overrightarrow{I_{s}} & = & \overrightarrow{I_{s}}^{pump}-\frac{g_{R}}{4\pi}\:\overrightarrow{m}\times\overrightarrow{V_{N}}\left(0\right)\times\overrightarrow{m}\nonumber \\
 &  & -\frac{g_{I}}{4\pi}\:\overrightarrow{V_{N}}\left(0\right)\times\overrightarrow{m}\label{eq:noncol}
\end{align}
where $g_{R}$ and $g_{I}$ are the real and imaginary spin mixing
conductances, $\overrightarrow{V_{N}}\left(0\right)$ is the spin
voltage vector at the interface on the N side, $\overrightarrow{m}\left(t\right)$
is the magnetization unit vector in the ferromagnet and the pumping
spin current\citep{wang_voltage_2006} is 
\begin{equation}
\overrightarrow{I_{s}}^{pump}=\frac{g_{R}}{4\pi}\:\overrightarrow{m}\times\dot{\overrightarrow{m}}+\frac{g_{I}}{4\pi}\:\dot{\overrightarrow{m}}.
\end{equation}

The internal spin conductance tensor can be directly read off Eq.
(\ref{eq:noncol}) : 
\begin{equation}
g_{IS}^{ij}=\frac{g_{R}}{4\pi}\:\left(\delta_{ij}-m_{i}\;m_{j}\right)+\frac{g_{I}}{4\pi}\;\varepsilon_{ijk}\;m_{k}.
\end{equation}
For circular precession of the ferromagnet around a direction $z$
with angle $\theta$ , $m_{z}=\cos\theta$ and if $g_{I}$ is negligible
(which is often the case), 
\begin{equation}
g_{IS}^{zz}=\frac{g_{R}}{4\pi}\;\sin^{2}\theta
\end{equation}
will be the dominant element of the tensor and one can use the internal
spin resistance relation for collinear batteries. The internal spin
resistance is then:
\begin{equation}
r_{IS}=\frac{4\pi}{g_{R}\:\sin^{\text{2}}\theta},
\end{equation}
while the maximum spin current is:
\begin{equation}
I_{s,max}=\frac{g_{R}}{4\pi}\;\frac{\hbar}{e}\;\omega\;\sin^{2}\theta
\end{equation}
 and the maximum spin voltage is just 
\begin{equation}
V_{s,max}=\hbar\omega/e
\end{equation}
. 

(Our result differs from the maximum spin voltage quoted in Ref. \citep{brataas_spin_2002}
as 
\begin{equation}
V_{s,max}=\hbar\omega\:\sin^{2}\theta/e\;\left(\sin^{2}\theta+\eta\right)
\end{equation}
 where $\eta$ is a positive load-dependent parameter. The discrepancy
is reconciled by observing that in our internal spin resistance definition,
the maximum voltage should be derived by finding the maximum against
any load while $\eta$ is a load dependent parameter. The max voltage
corresponds obviously to $\eta=0$ which recovers our value.)

The internal spin resistance is therefore in the range of $g_{R}^{-1}$
and can be tuned by varying the precession angle $\theta$; the spin
mixing conductance per unit area is typically in the $1\:f\Omega^{-1}/m^{2}$
(see for instance Ref. \citep{brataas_non-collinear_2006}) so for
an area $A=10^{-15}\:m^{2}$this means $r_{IS}$ in the $10-100\:\Omega$
range for precession angles between $0.1-1\:rad$. This implies that
with loads having spin resistance in the $1\:\Omega$ range, such
spin batteries will be good spin current source rather than good spin
accumulation sources. The load is a sufficiently 'bad spin sink' whenever
its spin resistance is larger than $r_{IS}$ (this is the 'spin battery'
regime discussed in Ref. \citep{brataas_spin_2002}). Observe also
that the smaller the area $A$, the larger $r_{IS}$ will be. This
implies that one will get better spin current sources, closer to ideality.

\subsection{Spin Hall effect\label{subsec:Spin-Hall-effect}}

\begin{figure}
\noindent \centering{}\includegraphics[width=0.8\columnwidth]{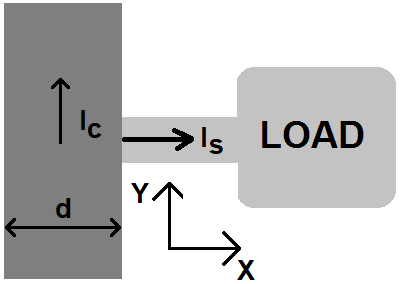}\caption{\label{fig:spinHall}A charge current $I_{c}$ flows in direction
$y$ in a bar of width $d$. As a result of spin orbit interaction
a transverse spin current $I_{s}$ is generated and feeds a load.}
\end{figure}

The spin Hall effect consists in the generation of a transverse spin
current from a charge current\citep{dyakonov_pis_1971,hirsch_spin_1999}.
We model the spin battery as a bar in direction $y$ of width $d$;
the load is perpendicular to the bar (parallel to direction $x$)
in order to collect the spin current at $x=0$ (see Fig. \ref{fig:spinHall}).
A spin accumulation is generated in the load which spills back into
the bar and decays with a relaxation spin length $l_{N}$. As a result
the spin current in the bar will have two components (in direction
$x$):
\begin{equation}
j_{s}=j_{s}^{SHE}+j_{s}^{SD}
\end{equation}
 where $j_{s}^{SD}$ obeys the spin diffusion equation while $j_{s}^{SHE}$
is the spin Hall spin current which does not depend on $x$.

The boundary condition for the spin current is $j_{s}\left(x=-d\right)=0$.
Using the diffusion equation 
\[
j_{s}^{SD}=+\frac{\sigma_{N}}{e}\:\partial_{x}\Delta\mu
\]
which relates spin accumulation and spin current one finds after straightforward
calculations:
\begin{align}
r_{IS} & =r_{N}\:\coth(d/l_{N}),\\
V_{s,max} & =\frac{\cosh\left(d/l_{N}\right)-1}{\sinh(d/l_{N})}\:r_{N}A\:j_{s}^{SHE}\\
 & =\tanh\left(\frac{d}{2l_{N}}\right)\:r_{N}A\:j_{s}^{SHE}
\end{align}
where $r_{N}$ is the bar spin resistance ($r_{N}=\rho_{N}^{*}\:l_{N}/A$)
and $A$ is the cross-section. In the limit of infinite width ($d\rightarrow\infty$),
the internal spin resistance is equal to $r_{N}$ the spin resistance
of the bar; in the opposite limit ($d\ll l_{N}$), the internal spin
resistance is infinite and one has an ideal spin current source. An
interesting consequence is that the internal spin resistance can be
tuned from the $1\:\Omega$ range to much larger values by varying
the width $d$.

In the limit of infinite width ($d\rightarrow\infty$) one finds 
\[
V_{s,max}=r_{N}A\:j_{s}^{SHE};
\]
in the opposite limit, 
\[
V_{s,max}=0
\]
 So one needs a wide bar, that is $d\gg l_{N}$ where $r_{IS}=r_{N}$.
The spin voltage can be detected through the ISHE. The magnitude is
up to several hundreds of microVolts.

\section{Conclusion\label{sec:Conclusion.}.}

We have introduced the concept of internal spin resistance which allows
to quantify whether a spin battery will be a good spin current source
(internal spin resistance much larger than the load spin resistance)
or a good spin accumulation source (opposite limit). This gives a
criterion for the design of spin batteries according to the main usage
targetted : angular momentum transfer or spin accumulation. The internal
spin resistance has been shown to be a positive number to ensure dissipation
(or more generally the internal spin tensor is positive semi-definite).
We have also considered several spin batteries relying on spin injection,
spin pumping or spin Hall effect, in collinear as well as non-collinear
settings, deriving explicit expressions for the internal spin resistance.

\appendix
\bibliographystyle{apsrev4-1}
\bibliography{spinemf}

\end{document}